\documentclass[
aps,
prl,
reprint,
superscriptaddress,
amsmath,
amssymb,
showpacs
]{revtex4-1}

\usepackage{graphicx}
\usepackage{multirow}
\usepackage{color}

\begin{document}


\title{Generation of High Brightness Electron Beams via Ionization Induced Injection by Transverse Colliding Lasers in a Beam-Driven Plasma Wakefield Accelerator}


\author{F. Li}
\author{J. F. Hua}
\author{X. L. Xu}
\author{C. J. Zhang}
\author{L. X. Yan}
\author{Y. C. Du}
\author{W. H. Huang}
\author{H. B. Chen}
\author{C. X. Tang}
\affiliation{Key Laboratory of Particle and Radiation Imaging of Ministry of Education, Tsinghua University, Beijing 100084, China}

\author{W. Lu}
\email{Email: weilu@tsinghua.edu.cn}
\affiliation{Key Laboratory of Particle and Radiation Imaging of Ministry of Education, Tsinghua University, Beijing 100084, China}
\affiliation{University of California Los Angeles, Los Angeles, California 90095, USA}

\author{C. Joshi}
\author{W. B. Mori}
\affiliation{University of California Los Angeles, Los Angeles, California 90095, USA}

\author{Y. Q. Gu}
\affiliation{Laser Fusion Research Center, China Academy of Engineering Physics, Mianyang, Sichuan 621900, China}


\date{\today}

\begin{abstract}

The production of ultra-bright electron bunches  using ionization injection triggered by two transversely colliding laser pulses inside a beam-driven plasma wake is examined via three-dimensional (3D) particle-in-cell (PIC) simulations. The relatively low intensity lasers are polarized along the wake axis and overlap with the wake for a very short time. The result is that the residual momentum of the ionized electrons in the transverse plane of the wake is much reduced and the injection is localized along the propagation axis of the wake. This minimizes both the initial ¡°thermal¡± emittance and the emittance growth due to transverse phase mixing.  3D PIC simulations show that ultra-short ($\sim$8 fs) high-current (0.4 kA) electron bunches with a normalized emittance of 8.5 and 6 nm in the two planes respectively and a brightness greater than $1.7\times10^{19} {\rm A\cdot rad^{-2}\cdot m^{-2}}$ can be obtained for realistic parameters.

\end{abstract}


\maketitle


The demonstration of the Linac Coherent Light Source (LCLS) as an X-ray free electron laser (X-FEL) \cite{emma2010} has given impetus to research on the ¡°fifth-generation¡± light sources \cite{pellegrini2010}. The goal is to make X-FELs smaller and cheaper while decreasing their wavelength  and increasing their coherence  and intensity. The FEL performance is partially determined by the brightness of the electron beam that traverses the undulator. The  brightness is defined as $B_n=2I/\epsilon_n^2$ where $I$ is the beam current and $\epsilon_n$ is the normalized emittance of the beam. In order to make the length of the undulator needed to drive the SASE-FEL \cite{hogan1998} into saturation, shorter, high current ($\sim$kA), multi GeV electron beams with $\epsilon_n \sim 10nm$  will be needed. These  emittances are an order of magnitude smaller than those from state-of-the-art photoinjector RF guns \cite{limborg2012}. In this letter, we show the generation of ultra-bright electron bunches using ionization injection triggered by two transversely overlapping laser pulses inside a beam-driven wake in plasma. In our scheme, the relatively low intensity lasers are polarized along the wake axis and overlap with the wake for a very short time. Particle-in-cell (PIC) simulations using OSIRIS  \cite{fonseca2002} show that this geometry reduces the residual momentum of the ionized electrons in the transverse plane and localizes them along the propagation axis of the wake leading to an electron beam with a brightness greater than $10^{19}$ A$\cdot$rad$^{-2}\cdot$m$^{-2}$ that would be highly attractive to future light sources.

When a dense ($n_b>n_p$, $k_p\sigma_{r,z}<1$), ultra-relativistic ($\gamma\gg1$) electron beam propagates through a plasma, the plasma electrons can be completely blown out by the repulsive Coulomb force of the  beam leaving behind a cavity of more massive ions \cite{rosenzweig1988,lu2006, lu2006pop} which then pull the electrons back creating a wakefield with a phase velocity equal to the velocity of the beam. Here $n_b$, $n_p$, $k_p$ and $\sigma_{r,z}$ are beam density, plasma density, inverse of the plasma skin-depth and transverse and longitudinal r.m.s. size of the electron beam respectively.  The accelerating and focusing fields inside this wakefield have ideal properties for acceleration of electrons while maintaining beam quality \cite{rosenzweig1988,lu2006, lu2006pop}  and high-gradient acceleration by such wakes has been experimentally demonstrated \cite{joshi2002,muggli2004,hogan2005,blumenfeld2007}.

For a plasma density $\sim10^{18}\;{\rm cm^{-3}}$, the wavelength of the ion cavity is about several tens of microns making synchronization and efficient capture of externally injected electrons into such a cavity extremely challenging. Self-injection of electrons in plasma wakes  is conceptually simple, however, it still can not generate sufficiently high brightness beams needed for next generation light sources\cite{plateau2012,wiggins2010}. Other electron injection schemes such as ponderomotive force injection have been proposed \cite{umstadter1996} for laser-plasma wakefield accelerators. Here a pump laser pulse excites the wake and a second injector pulse propagates orthogonally to the pump disrupting the orbits of some of the plasma electrons leading to trapping. Soon thereafter the collinear colliding pulse injection was suggested \cite{esarey1997} and experimentally demonstrated \cite{faure2006}. In addition a sudden \cite{suk2001} or gradual \cite{geddes2008} density transition from a high plasma density to a low plasma density has also been shown to inject particles into plasma wakes. Another technique is ionization injection where electrons are produced inside the wake by the electric field of a laser pulse or the drive electron beam (that produce the wake) where they can be more easily captured and accelerated. Ionization injection is attractive because it offers the potential to control the charge and emittance of the accelerated beam. Very recently it was proposed to combine the ionization injection via an auxiliary laser pulse into a beam driven wake \cite{hidding2012}. This approach allows the use of a lower intensity ionizing laser, thereby further reducing residual momentum and hence the transverse emittance of the injected electrons. In this letter, we show that electron injection into a beam-driven PWFA via tunnel ionization in the overlap region of two laser pulses (moving transversely across the wake) can generate an electron beam with extremely small transverse emittances and therefore an extremely high brightness.

This mechanism is explored using the 3D PIC code OSIRIS \cite{fonseca2002} in Cartesian coordinates using a moving window. We define the $z$-axis to be the propagating direction of the drive beam, and the $x$-axis to be the propagating direction of the colliding laser pulses with their electric field polarized along the $z$-axis. The simulation window had a dimension of $89\times81\times121$ $\mu$m with $1400\times512\times760$ cells in the $x,y$ and $z$ directions respectively.  This corresponds to cell sizes of $0.5k_0^{-1}$ in the $x$ direction and $1.25k_0^{-1}$ in the $y$ and $z$ directions. The code uses Ammosov-Delone-Krainov (ADK) tunneling ionization model \cite{ammosov1986}.

For simplicity, the simulation is initialized with plasma with a density of $n_p=2.4\times10^{17}$ cm$^{-3}$, represented by 8 particles per cell, and neutral He with a density of $1.1\times10^{18}$ cm$^{-3}$ represented by 8 neutral atoms per cell. The pre-ionized plasma can be viewed as a fully ionized separate gas.  A 500 MeV drive beam  of the form
 $\frac{N}{(2\pi)^{3/2}\sigma_z\sigma_r^2} e^{-r^2/2\sigma_r^2} e^{-z^2/2\sigma_z^2}$with $\sigma_z=11.4$ $\mu$m, $\sigma_r=7.6$ $\mu$m, and $N=1.25\times10^9$ (200pC) respectively, propagates through the plasma and excites the wake. The self electric field of the beam ($\sim$50 GeV/m) does not ionize the helium atoms. In addition, two counter propagating laser pulses moving along the + and - $x$-axis directions are synchronized with the electron beam so that they overlap inside the first bucket (ion cavity) near the point where the longitudinal electric field vanishes, \emph{i.e.}, where $E_z=0$. Each laser has a normalized vector potential of $a_0=0.016$, a duration of $\tau=20$ fs, and a focal spot size of $w_0=6$ $\mu$m. These parameters correspond to each laser having a focused intensity of $5.5\times10^{14}$ W/cm$^2$ for a wavelength of 800 nm.

\begin{figure}
    \includegraphics[width=8.6cm]{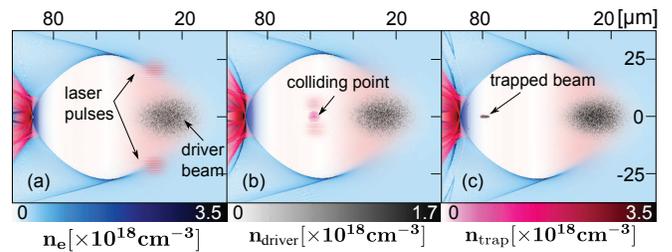}
    \caption{\label{fig:snapshots}Snapshots from PIC code simulations illustrating the transverse colliding pulse injection of helium electrons into the ion cavity. Snapshots (a) to (c) show the charge density distribution of driver beam, wake electrons and helium electrons at three different times (a) $\sim$80 fs before laser pulses collision (b) around laser pulses' collision time (c) $\sim$200 fs after collision when the injected electrons become trapped in the wake. The ionized electrons seen on the left of each figure are due to the wakefields themselves exceeding the ionization threshold for helium at the rear of the first bubble however none of these electrons are trapped and accelerated in the wakefield.}
\end{figure}

We first examine the injection process. It is easier to trap and to control the self-injection of an electron that is born (ionized) at rest inside the wake. The trapping threshold can be written as \cite{pak2010} $\Delta \psi \equiv \psi-\psi_\text{init} < -1 + (1+[p_\perp/mc]^2)^{1/2} / \gamma_{ph}$ where $\psi \equiv e(\phi-A_z)/mc^2$ is the normalized wake potential and $\psi_\text{init}$ is the wake potential when and where an electron is created. The wake potential is at a minimum at the rear of the cavity so it is easiest to trap an electron if it is created when the potential is at a maximum (a zero for $E_z$) which occurs in the middle of the cavity. Furthermore,  low emittance beams are generated if the electrons are born near the transverse axis. For laser ionization,  an electron is born inside a laser so it also acquires a residual drift  in the polarization direction of the laser of $p/mc=eA_\text{init}/mc^2$ (approximately in multi-dimensions). The amount of charge that will be injected is controlled by the neutral gas density.

Fig. \ref{fig:snapshots} illustrates the injection process through a series of snapshots that shows the formation of the wake, the collision of the two lasers triggering injection within this wake through ionization, and the trapping of the injected electrons, where the densities of the drive electron beam, wakefield electrons and laser-ionized electrons are plotted with colors black, blue and pink, respectively. In each snapshot a stable highly nonlinear wakefield cavity is seen. In Fig. \ref{fig:snapshots}(a) the injector laser pulses are seen just moving across the sharp electron sheath of the ion cavity. Because of the low intensity of laser pulses, approximately $5.5\times10^{14}$ W/cm$^2$, the ponderomotive force of the laser pulse is not enough to perturb the sheath electrons, and thus the wake structure remains unaffected. In Fig. \ref{fig:snapshots}(b), the lasers collide on the axis, where they have the maximum (overlapping) intensity at the position where $E_z=0$. The local laser intensity exceeds the ionization threshold only where the lasers overlap (each laser has an intensity below the ionization threshold) and a large fraction of neutral helium atoms within this volume is now ionized. As the laser pulses travel past the collision point the injection ceases. Note that the drift momentum of the electrons is along the laser polarization direction and therefore now affects predominantly the longitudinal momentum spread of the beam, leading to a longitudinal emittance of 0.06 keV$\cdot$ps in this example. These laser ionized helium electrons then respond to the wake fields and are rapidly accelerated to a longitudinal velocity close to the phase velocity of wakefield by the time they have slipped backwards to the rear of the ion cavity. They then begin to move synchronously with the wake, as depicted in Fig. \ref{fig:snapshots}(c).

\begin{figure}
    \includegraphics[width=8.6cm]{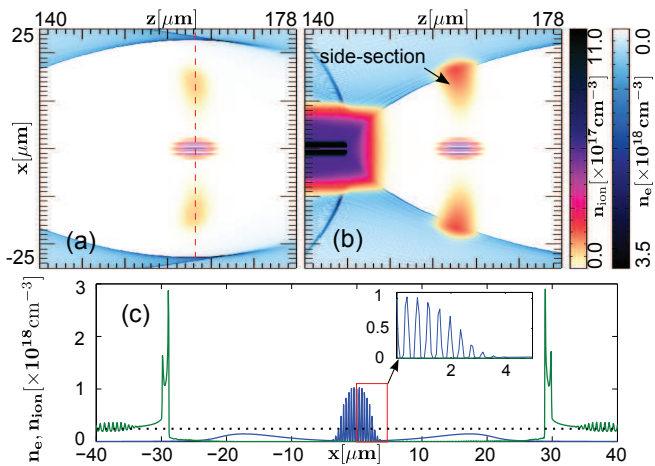}
    \caption{\label{fig:ion}Ion density (color-coded) indicates the ionization level of helium gas. Wake electrons are colored blue. (a) the ionization level at the instant of collision. (b) 80 fs after the collision, when the laser pulses are on the verge of exiting the electron sheath. (c) The lineout on the red dashed line in (a). Laser-ionized He$^+$ is shown in blue, sheath electron density is shown in green and black dotted line indicates the initial ion background.}
\end{figure}

The details of the ionization process can be more clearly seen by plotting the density of the He$^+$ ions. The superposition of the two  lasers  gives rise to a standing wave  with a wavelength $\lambda_0/2$,  a frequency $\omega_0$, and an amplitude four times that of a single pulse. Fig. \ref{fig:ion} shows the He$^+$ ion density distribution resulting from laser ionization at two different times. The laser created He$^+$ ions are immobile because they are more massive, thus their location represents the birthplace of the electrons. In Fig. \ref{fig:ion}(a), one can observe the layered ion distribution reflecting the standing wave of electric field of the two laser pulses at the instant where the lasers overlap. In Fig. \ref{fig:ion}(c) a lineout corresponding to the red dashed line of Fig. \ref{fig:ion}(a) is shown. The inset shows that under such laser intensity, He atoms are almost fully ionized to He$^+$ at the antinodes of the overlapping intensity. Helium electrons are mainly born within 2 $\mu$m of the axis. In Fig. \ref{fig:ion}(b) we show a snapshot of ion density 80 fs after the laser  collision at which time the laser is near the sheath. The superposition of the laser and wake fields near the electron sheath leads to some off-axis ionization. Additionally, there is ionization from the wakefield alone near the rear of the bubble. Fortunately, in this case these off-axis electrons (and those in the rear) do not get trapped because their initial positions are too close to the rear   and too far from the axis of the cavity. However, for other beam and plasma parameters, these undesired He electrons can get trapped  deteriorating the beam quality.

Fig. 3 illustrates the phase space distributions in each plane about 500 fs after the collision. The projected normalized emittance is obtained by $\epsilon_n=\overline{\beta\gamma}\sqrt{\langle x'^2\rangle\langle x^2\rangle-\langle xx'\rangle^2}$ for each of the two transverse planes. Initially, the injected beam has an  ultra-low projected transverse normalized emittance for the whole bunch, of about 8.5 nm in the $x$ direction and 6 nm in the $y$ direction, which  is observed to be invariant after propagating 100 $\mu$m. At this distance the beam has an average energy 5.3 MeV, a slice energy spread of $\sim$12 keV and a total charge of 4.6 pC. The projected longitudinal emittance as explained earlier is also very small, about 0.06 keV$\cdot$ps. The beam current profile is near flattop, with a r.m.s. pulse duration around 8 fs and peak current 0.44 kA. The brightness of the beam $B_n=2I_p/\epsilon_n^2$ is estimated to be $1.7\times10^{19} {\rm A\cdot rad^{-2}\cdot m^{-2}}$, more than 3 orders of magnitude higher than that of LCLS. In principle, the trapped beam charge and the normalized emittance can be further optimized by adjusting the neutral gas concentration and the spatial sizes of injector pulses.

\begin{figure}
    \includegraphics[width=8.6cm]{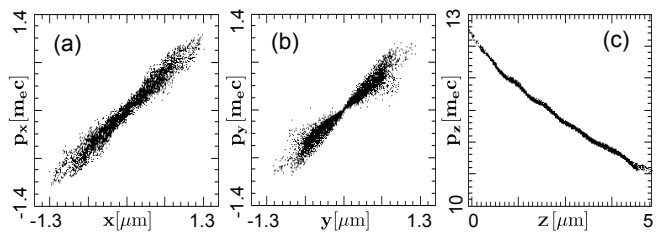}
    \caption{\label{fig:trans_phsp}The (a) $x$-$p_x$, (b) $y$-$p_y$ and (c) $z$-$p_z$ phase space distribution about 500 fs after the pulse collision.}
\end{figure}

To achieve the extremely small $\epsilon_n$, certain conditions have to be fulfilled. First, the electrons have to "start" with small transverse momenta. In the proposed scheme, since the laser pulses are polarized along $z$ axis and propagate perpendicularly to the x axis, the electric (and vector) field components $E_x,E_y,E_z$ scale as $E_x\sim\epsilon^2E_z,\,E_y\sim\epsilon E_z $, where $\epsilon$ is a characteristic small parameter defined as $\epsilon\equiv1/k_0w_0$, around 0.02 in our simulation, therefore the residual transverse momenta $p_x,p_y$ of electrons just after ionization scale as $p_x/m_ec\sim\epsilon^2a_0,\,p_y/m_ec\sim\epsilon a_0$, which are extremely small. Second, the transverse electron beam sizes $w_b$ are determined by the laser intensity contour above ionization threshold near collision point, approximately $w_b\propto c\tau$, which for the simulated parameters is around a few microns. Combining the above two factors, the intrinsic or "thermal" emittance of the electron beam, which is defined as $\epsilon_{\rm th}=\langle p_\perp\rangle w_b $, is very small.

However, in order to preserve the "thermal" emittance, it is necessary to avoid ¡°phase mixing ¡±, which arises when electrons within the bunch are born at different times and therefore at different phases of their betatron oscillations \cite{xu2012}. To illustrate this point, using OSIRIS we simulate the injection process of both the transverse colliding pulse injection scheme (this paper) and the recently proposed longitudinal injection scheme \cite{hidding2012}. In the longitudinal injection scheme, a single injection laser pulse propagates colinearly at an optimum distance (67 fs) behind the beam driver, also at the phase where $E_z =0$. For both simulations, the drive beam is the same, with 200 pC of charge and  $\sigma_r=\sigma_z=3.8$ $\mu$m, and the pre-ionized plasma density is set to $5\times10^{17}$ cm$^{-3}$.The simulation domain is $63.5\mu{\rm m}\times50.8\mu{\rm m}\times63.5\mu{\rm m}$ and $1000\times400\times500$ cells were used in the transverse injection case and $500\times400\times1000$ cells in the longitudinal injection case. The injection lasers have the same pulse duration ($\tau=20$ fs) and a FWHM spot size (5 $\mu$m), but with different focal intensities. For longitudinal injection, the laser is polarized in $x$ direction and focused with $a_0=0.035$ or an intensity of $2.6\times10^{15}$ W/cm$^2$.  For  the  colliding pulse injection, each laser has a $a_0=0.016$, therefore near the antinode of the  standing wave, the combined laser intensity is just above the ionization threshold of helium. The neutral helium densities are also different. The He density is $1.1\times10^{18}$ cm$^{-3}$, and $5.2\times10^{16}$ cm$^{-3}$ for the "transverse" and "longitudinal" propagation cases respectively. By doing so, both cases obtain similar injected charge (1.9 pC for transverse injection and 3 pC for longitudinal injection). If we choose much higher helium density in the longitudinal injection case, the injected charge can be larger but the emittance gets much worse, e.g.,  for the same He density as that used in the transverse injection case $\epsilon_n$ was  ten times worse. We also have simulated the colliding pulse geometry with the lasers polarized along y instead of x. In this case the thermal emittance along y is comparable to that of the longitudinal case illustrating the importance of minimizing both the thermal emittance and phase mixing.

\begin{figure}
    \includegraphics[width=8.6cm]{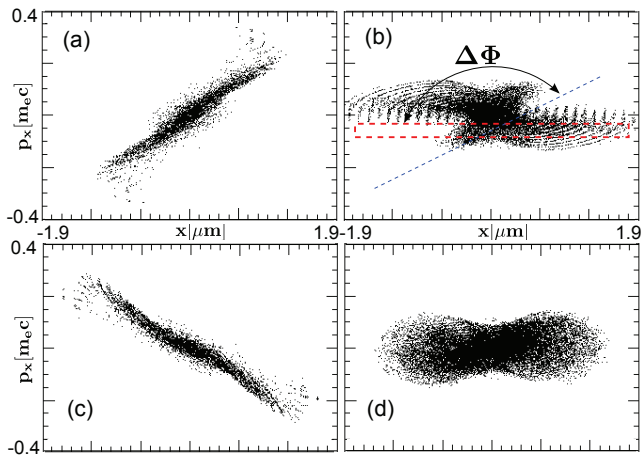}
    \caption{\label{fig:phspcomp}Comparison of $x$-$p_x$ phase space evolution between the transverse colliding pulses injection (left column) and the longitudinal injection (right column) at the instant 130 fs (top) and 260 fs (bottom) after the onset of injection. Phase-mixing occurs during the longitudinal injection.}
\end{figure}

Fig. \ref{fig:phspcomp} illustrates that the $x$-$p_x$ phase space evolution is very different between the transverse colliding pulses injection (left column) and the longitudinal injection (right column). Fig. \ref{fig:phspcomp}(a) and (b) are taken 130 fs after the onset of injection (at this time  injection has ceased  for colliding pulses  while it is still ongoing for the longitudinal injection scheme). Fig. \ref{fig:phspcomp}(c) and (d) are taken about 260 fs after the onset of injection. There is phase space rotation in the colliding pulse case (as expected) and there is far less phase space mixing as compared to  longitudinal injection, because the injection distance in the longitudinal scheme is much longer than that in the colliding pulse scheme. The injection distance in the longitudinal case is on the same order as the Rayleigh length of the laser pulse, $z_R=\pi w_0^2/\lambda_0 $. Over this distance He electrons born at different times have different Betatron motion \cite{wang2002} phase $\Phi$. We define  $\Delta\Phi$ as the betatron phase difference between the first and the last ionized electron. As seen in Fig. \ref{fig:phspcomp}(b), the first  ionized electrons have rotated to the blue dashed line, while the final ionized electrons have just been released (roughly in the red dashed box). Xu. \emph{et al}. have studied the electron dynamics of ionization injection and shown that the phase divergence $\Delta\Phi$ is mainly determined by injection distance $\Delta$ as $\Delta\Phi\propto\sqrt{\Delta}$ \cite{xu2012}. The implication being that a longer injection distance will cause more phase mixing leading to a larger final emittance. Therefore, the effect of phase-mixing is found to be small in the proposed scheme,  because all the He electrons are released in short time which is controlled by the overlap time of laser pulses. As shown in Fig. \ref{fig:phspcomp}(a) and (c), the electrons rotate in the phase space with a small phase divergence, similar to a laminar beam.  In our simulations, the normalized emittances in the longitudinal injection scheme are 34 nm in $x$ direction and 25 nm in $y$ direction, both four times larger than those obtained in the transverse colliding pulses injection scheme. We note that if an energy chirp grows along the beam,  then different slices of the beam will oscillate at different betatron frequencies leading to a growth in the projected (not slice) emittance.

\begin{figure}
    \includegraphics[width=8.6cm]{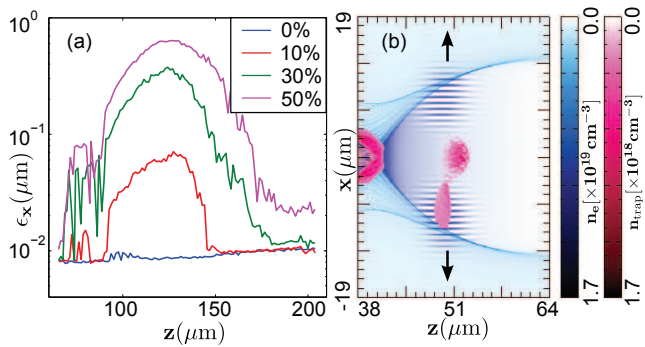}
    \caption{\label{fig:imbalance}The impact of laser power imbalance on the transverse emittance. (a) the transverse normalized emittance variation during the trapping process under different laser power mismatch ratios. (b) additional helium electrons generated on the higher power side for the 30\% power mismatch.}
\end{figure}

We have examined the effect of power imbalance between the two colliding laser pulses on the injection process. Assuming the total power of the two pulses is fixed, we define the power mismatch ratio $R=1-P_L/P_H$, where $P_L, P_H$ represent the lower and higher laser power respectively. We use simulation parameters identical to that of Fig. \ref{fig:phspcomp}(a)  and the results are shown in Fig. \ref{fig:imbalance}.

In Fig. \ref{fig:imbalance}(a), the evolution of  $\epsilon_{n,x}$ starting from the birth of helium electrons is plotted. Except for the balanced case of $R=1$, all other curves rise rapidly first and then eventually drop to a low level. This reduction can be understood from the electron density plot shown in Fig. \ref{fig:imbalance}(b). After the collision is over, a large number of additional helium electrons are liberated in the wake by the overlapping of the higher power laser and the wake fields.  These electrons lead to the emittance growth because they are born with transverse positions far off axis. However, the majority of these additional electrons do not get trapped ($\Delta\Psi>-1$) and fall behind leading to the drop in the emittance over time. The final emittance  can be as low as 10 nm as long as the power mismatch is less than 30\% and with approximately the same charge. We note that there is much flexibility in this idea as one could experiment with different gas mixtures and colliding pulse geometries to optimize the tradeoffs in charge and emittance.

In conclusion, we have proposed a new injection scheme for PWFA by utilizing transverse colliding laser pulses. This scheme can generate extremely low emittance electron beams with sufficiently high current. Using simulations, we demonstrate the possibility of generating beams with $I_p=0.44$ kA (4.6 pC of charge), $\epsilon_{n,x}=8.5$ nm, $\epsilon_{n,y}=6$ nm and longitudinal emittance of 0.06 keV$\cdot$ps leading to a beam brightness greater than $1.7\times10^{19}$ A$\cdot$rad$^{-2}\cdot$m$^{-2}$, which is three orders of magnitude higher than that can be obtained at the LCLS, therefore this scheme may have important impact on future FEL research.

The work at Tsinghua University was supported by NSFC grants 11175102 and 11005063. The work at UCLA was supported by DOE grant DE-FG02-92-ER40727, NSF grant PHY-0936266. The computations were carried out on the Hoffman and Dawson2 Clusters at UCLA.

\bibliography{ref}

\end{document}